\documentclass[baaa]{baaa}

\usepackage[pdftex]{hyperref}
\usepackage{subfigure}
\usepackage{natbib}
\usepackage{helvet,soul}
\usepackage[font=small]{caption}

\begin{document}


\journalvol{61A}
\journalyear{2019}
\journaleditors{R. Gamen, N. Padilla, C. Parisi, F. Iglesias \& M. Sgr\'o}


\contriblanguage{1}


\contribtype{2}

\thematicarea{7}

\title{Formation of massive black holes via collisions and accretion}


\titlerunning{Formation of massive black holes}


\author{D.R.G. Schleicher\inst{1}, M.~A. Fellhauer\inst{1}, T.~Boekholt\inst{2}, B.~Reinoso\inst{1}, R.S. Klessen\inst{3,4}, M.Z.C. Vergara\inst{1},  P.J. Alister Seguel  \inst{1}, S.~Bovino\inst{1}, C.~Olave\inst{1}, V.B.~D\'iaz\inst{1},  FFibla \inst{1}, R. Riaz\inst{1}, B. Bandyopadhyay\inst{1}, R.I.~San Martin-Perez\inst{1}, J.~Zamponi\inst{1}, L.~Haemmerle\inst{5}  }
\authorrunning{Schleicher et al.}


\contact{bidisharia@gmail.com}

\institute{
Departamento de Astronom\'ia, Facultad Ciencias F\'isicas y Matem\'aticas, Universidad de Concepci\'on, Av. Esteban Iturra s/n Barrio Universitario, Casilla 160-C, Concepci\'on, Chile \and
CIDMA, Departamento de F\'isica, Universidade de Aveiro, Campus de Santiago, 3810-193 Aveiro, Portugal \and
Universit\"at Heidelberg, Zentrum f\"ur Astronomie, Institut f\"ur Theoretische Astrophysik, Albert-Ueberle-Str. 2, 69120 Heidelberg, Germany \and
Universit\"at Heidelberg, Interdisziplin\"ares Zentrum f\"ur Wissenschaftliches Rechnen, Im Neuenheimer Feld 205, 69120 Heidelberg, Germany \and
Observatoire de Gen\`eve, Universit\'e de Gen\`eve, chemin des Maillettes 51, CH-1290 Sauverny, Switzerland
}


\resumen{
Para explicar la poblaci\'on de agujeros negros supermasivos en $z\sim7$, se necesitan semillas de agueros negros muy masivas o, alternativamente, escenarios de acreci\'on super-Eddington para alcanzar masas finales del orden de $10^9$~M$_\odot$. El modelo del colapso directo predice la formaci\'on de un solo objeto masivo debido al colapso directo de una nube de gas masiva. Las simulaciones en los \'ultimos a\~nos han demostrado que este escenario es muy dif\'icil de lograr. Un modelo realista de formaci\'on de agujeros negros deber\'ia tener en cuenta la fragmentaci\'on y considerar la interacci\'on entre los procesos estelar-din\'amicos y la din\'amica del gas. Presentamos aqu\'i una simulaci\'on num\'erica aplicada con el c\'odigo AMUSE, realizando un tratamiento aproximado del gas. Seg\'un estas simulaciones, mostramos que se pueden formar agujeros negros muy masivos de $10^4-10^5$~M$_\odot$ dependiendo del sumistro del gas y la acreci\'on en las protoestrellas.
}

\abstract{
To explain the observed population of supermassive black holes at $z\sim7$, very massive seed black holes or, alternatively, super-Eddington scenarios are needed to reach final masses of the order of $10^9$~M$_\odot$. A popular explanation for massive seeds has been the direct collapse model, which predicts the formation of a single massive object due to the direct collapse of a massive gas cloud. Simulations over the last years have however shown that such a scenario is very difficult to achieve. A realistic model of black hole formation should therefore take fragmentation into account, and consider the interaction between stellar-dynamical and gas-dynamical processes. We present here numerical simulations pursued with the AMUSE code, employing an approximate treatment of the gas. Based on these simulations, we show that very massive black holes of $10^4-10^5$~M$_\odot$ may form depending on the gas supply and the accretion onto the protostars.
}


\keywords{black hole physics - stars: Population III - methods: numerical}

\maketitle

\section{Introduction}
At present, more than 100 quasars are known at $z>5.6$ \citep{Banados16, Schleicher18}, with the currently known highest-redshift quasar at $z\sim7.5$, hosting a supermassive black hole with about 800 million solar masses \citep{Banados18}. To explain such early and supermassive black holes, one needs to assume almost continuous Eddington accretion since redshifts $z\sim20$, bursts of Super-Eddington accretion or rather massive seeds at the beginning of the accretion process \citep[e.g.][]{Shapiro05}. The pathways to produce massive seeds have already been laid out by \citet{Rees84}, and include the formation via the direct collapse of a massive gas cloud into one single object, but also black hole formation via stellar-dynamical processes as a result of runaway-collisions either in stellar clusters or clusters of stellar-mass black holes.

The direct collapse seemed initially very promising, due to its capacity of producing very high-mass seeds with up to $10^5$ solar masses \citep[e.g.][]{Bromm03, Wise08, Schleicher10, Latif13, Latif14}.  However, to reach the required conditions, the collapse should basically be isothermal under atomic cooling, requiring very large ambient UV fluxes \citep{Latif15}. On the other hand, even tiny amounts of dust can already trigger strong fragmentation through the enhanced cooling \citep[e.g.][]{Omukai08, Dopcke11, Klessen12, Bovino16, Latif16}.

Under realistic conditions, it is then almost unavoidable for fragmentation to occur, at least initially preventing the formation of a single massive object. However, also star clusters can produce massive black holes by collisions, as discussed e.g. by \citet{Devecchi12,Sakurai17}, showing that black holes of $\sim500$~M$_\odot$ are able to form. It has been established over the last years that the radii of primordial protostars can be considerably enhanced while accreting, thereby effectively behaving like red giants \citep{Hosokawa13, Haemmerle18}. As known already from present-day protostellar clusters \citep{Baumgardt11}, such an enhancement of the protostellar radii increases the probability for collisions, and may favor the formation of massive objects. In the context of primordial stellar clusters, \citet{Reinoso18} recently explored the implications of such enhanced radii for the collisions, showing that indeed both the fraction of collisions as well as the mass of the central massive object increases significantly with protostellar radii. They also provide scaling relations that allow to infer the black hole mass as a function of the ambient condition.

The presence of gas in the first proto-cluster may however have further implications. First, it provides an additional gravitational potential, which enhances the velocity dispersion of the embedded stellar cluster. In addition, in the presence of gas, the protostars may accrete and change their masses during the run-time of the simulations. The interaction between gas-dynamical and stellar-dynamical processes may thus be quite relevant for the formation of very massive objects in the first stellar clusters, but has hardly been considered so far. In the following, we present a set of simulations pursued with the publicly available AMUSE framework\footnote{Webpage AMUSE: http://amusecode.org/} \citep{AMUSE13} to approximately account for such effects. A more detailed description of the simulations has been presented by \citet{Boekholt18}.

\section{Numerical setup}
We adopt here a simplified initial condition, where protostars and the gas both follow a Plummer distribution \citep{Plummer11}. In our reference model, the initial gas mass corresponds to $10^5$~M$_\odot$, the Plummer radius $0.1$~pc, the initial number of stars is $256$, with very low masses of initially $0.1$~M$_\odot$. The system is thus initially gas-dominated. We also introduce a cut-off radius after which the density is set to zero. The latter corresponds to five times the Plummer radius. The gravitational interaction between the stars is modeled via the N-body code ph4 by \citet{McMillan96}, using a fourth-order Hermite algorithm employing the time-symmetric integration scheme developed by \citet{Hut95}. The gravitational potential of the gas cloud is described as an analytic background potential, which is coupled to the stars using the  \texttt{BRIDGE} method \citep{bridge07}.

The accretion of the gas onto the stars is described by the simplified models outlined in Table~1 \citep{Boekholt18}. Models with an infinite gas reservoir indicate that the gas is efficiently resupplied during the evolution, so that gas accreted onto the protostars is not removed from the gas (models 1-2), while it is removed in the models with a finite gas reservoir (models 3-6). In case of position-dependent accretion, we assume that the accretion rate is proportional to the gas density in the Plummer sphere. If the accretion rate is also time-dependent, we assume it to be proportional to the mass in the gas reservoir. In models 3-6, the accretion is switched off when the gas reservoir is exhausted. Our fiducial accretion rate is $0.03$~M$_\odot$~yr$^{-1}$, as suggested through numerical simulations \citep[e.g.][]{Latif13, Latif14}, but we also explored other values.

The stellar radii have been determined using approximate fits to the mass-radius relations given by \citet{Hosokawa13} and \citet{Haemmerle18}, where both prescriptions yield similar results.  To model collisions, we adopt the so-called ''sticky-sphere'' approximation, replacing two protostars by one if the distance between two protostars becomes less than their radii. During the collision, we assume the conservation of mass, and the new radius is determined from the mass-radius parametrization. Using this setup, we follow the evolution of the system until no further collisions occur. The latter usually corresponds to a physical time of less than one million years.

\begin{table}
\centering
\begin{tabular}{llll}
\hline
Model & Gas reservoir & Position dep. & Time dep. \\
      &               & accretion    & accretion\\
\hline
1 & Infinite & no  & no \\
2 & Infinite & yes & no \\
3 & Finite   & no  & no \\
4 & Finite   & yes & no \\
5 & Finite   & no  & yes \\
6 & Finite   & yes & yes \\
\hline
\end{tabular}
\caption{ The gas accretion models for Pop.~III protostars embedded in their natal gas cloud, as presented by \citet{Boekholt18}.  }
\label{tab:accretion_models}
\end{table}

\section{Results}
The results for our fiducial parameters, providing the time evolution of the mass of the most massive central object, are presented in Fig.~1 for the different accretion models \citep{Boekholt18}. Clearly, the most optimistic models are models 1 and 2 with an infinite gas reservoir, so that the central mass can continue to grow unimpededly. Such a scenario is only feasible in case of a strong external mass supply, which needs to be transported through the protogalaxy for instance as a result of gravitational torques. In case the gas reservoir is however finite, the mass of the resulting object is lower by a  factor of 2 or 3. The most conservative model with respect to the black hole mass is our model 5, with the finite gas reservoir and a time-dependent accretion rate, which is however independent of position. In this case, large fractions of the mass are accreted in the outer parts of the cluster, where they do not strongly participate in collision events, and as a result they do not contribute to the mass of the most massive object. Nevertheless, even in the most conservative scenario, the central object reaches a mass of $10^4$~M$_\odot$.

We have checked the dependence of these results on various parameters, as presented in detail by \citet{Boekholt18}. We found that the most crucial parameter is the initial gas reservoir. In case a gas mass of $10^5$~M$_\odot$ is available, the results only weakly depend on other parameters, but the mass of the central massive object decreases significantly if less gas is available. The size of the cluster as well as the accretion rate, on the other hand, were found to be of minor relevance and to primarily affect the timescale until the process is completed. The model therefore has the advantage to potentially be robust, at least under conditions where a sufficiently large gas masses are available for accretion.

\section{Summary and discussion}

We find that our scenario provides a potentially promising pathway for the formation of very massive black holes, with masses between $10^4$ and $10^5$~M$_\odot$. Our gas-dynamical model is however still based on highly simplifying assumptions, considering an analytic gravitational potential and very simple prescriptions for the accretion of the protostars. For the future, it will be central to gradually relax some of the assumptions, and to explore how a detailed dynamical treatment of the gas and the accretion will affect the results. Additional uncertainties concern the geometry of the star and gas distribution, which was here assumed to be spherical. Numerical simulations however frequently suggest that fragmentation will occur preferentially in flattened geometries including rotation. The presence of such an ordered velocity component could potentially somewhat alter the results presented here, as collisions may become less likely in such a case. On the other hand, the presence of gas is also expected to lead to dynamical friction, which we neglected in the simulations presented here. Such friction could further affect the motions of the stars and increase the probability of collisions. 

As semi-analytic models indeed suggest that massive seeds are needed to form supermassive black holes \citep{Valiante16}, it will be important to explore realistic formation scenarios.

\begin{figure}
\centering
\begin{tabular}{c}
\includegraphics[height=0.38\textwidth,width=0.45\textwidth]{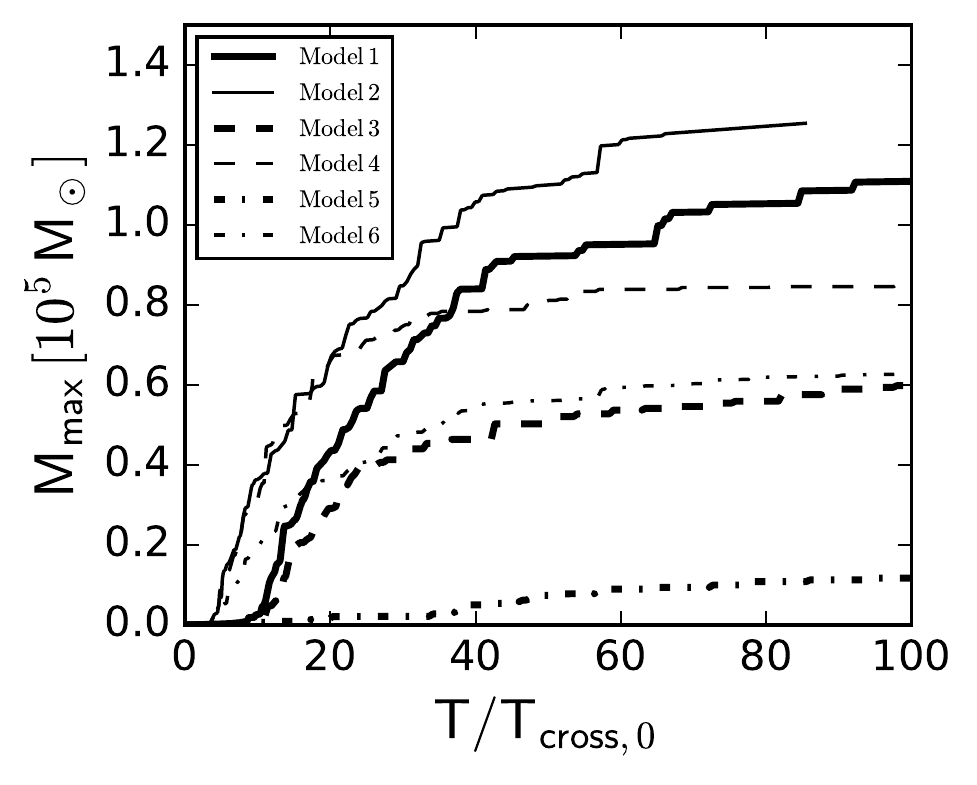} \\
\end{tabular} 
\caption{ Time evolution of the mass of the most massive object. Shown are the results for the six different accretion models given in Table~1. Figure from \citet{Boekholt18}. }
\label{fig:individual_Mmax}
\end{figure}

\begin{acknowledgement}
We thank for funding via Conicyt, in particular via Conicyt PIA ACT172033, Fondecyt regular (project code 1161247), the ''Concurso Proyectos Internacionales de Investigaci\'on, Convocatoria 2015'' (project code PII20150171) and the CONICYT project Basal AFB-170002. MF thanks for funding via Fondecyt regular 1180291. BR and VBD thank Conicyt for
financial support on their Master studies through CONICYT-PFCHA/Mag\'isterNacional/2017-22171385 and CONICYT-PFCHA/Mag\'isterNacional/2017-22171293. RSK acknowledges
financial support from the Deutsche Forschungsgemeinschaft via SFB 881, ``The Milky Way System'' (sub-projects B1, B2 and B8). He also thanks for support from the
European Research Council via the ERC Advanced Grant ``STARLIGHT: Formation of the First Stars'' (project number 339177).
TB acknowledges support from Funda\c{c}\~ ao para a Ci\^ encia e a Tecnologia (grant SFRH/BPD/122325/2016), and support from
Center for Research \& Development in Mathematics and Applications (CIDMA) (strategic project UID/MAT/04106/2013), and
from ENGAGE SKA, POCI-01-0145-FEDER-022217, funded by COMPETE 2020 and FCT, Portugal.
 \end{acknowledgement}
 
 \bibliographystyle{aa}
\small
\bibliography{bidibib}

 %

\end{document}